\documentclass[12pt]{article}
\usepackage{latexsym,epsfig}
\usepackage{hyperref}
\usepackage{color}

\begin{document}

%

~

\centerline{\bf \large What Exactly is Antimatter (Gravitationally Speaking)?}
%
%

\bigskip

\centerline{S. Menary}
\smallskip

\centerline{Dept. of Physics \& Astronomy}
\centerline{York University, 4700 Keele Street}
\centerline{Toronto, ON M3J 1P3, Canada}
\bigskip

\date{\today}
\begin{abstract}
There has been renewed interest in the idea of antigravity -- that matter and antimatter repel gravitationally - in lieu of the recent beautiful\footnote{I may be somewhat biased seeing as I'm a member of the ALPHA collaboration!} ALPHA-g result\cite{ALPHA-g} for the free-fall acceleration of antihydrogen of $a_{\bar{H}}=(0.75\pm 0.13~({\rm stat.+syst.})\pm 0.16~({\rm simulation}))g$. Precision tests of the Weak Equivalence Principle (WEP) have shown that binding energy (atomic, nuclear, and nucleonic) acts like matter under gravity. Whereas the contribution of atomic binding energy to the mass of antihydrogen is negligible, the majority of the mass of the antiproton comes from the gluonic binding energy. Hence, in terms of antigravity, the antiproton is mostly composed of matter and so even if the antimatter content of the antiproton is repelled by the Earth, there would still be a net attraction of the antihydrogen to the Earth because of its dominant matter content. Using recent Lattice QCD results showing that around two-thirds of the mass of the proton (and, hence, from the $CPT$ Invariance of QCD, of the antiproton) is due to gluons, I find that in the antigravity scenario the free-fall acceleration of antihydrogen would be $a_{\bar{H}}=(0.33^{+0.23}_{-0.11})g$. The fact that antinucleons are more matter than antimatter (in the gravitational sense) leads to quite different cosmological consequences than the naive antigravity scenario (where antistars are wholly antimatter). For example, it follows naturally that there is a matter-antimatter asymmetry in the universe right from the instant of the Big Bang -- but not a baryon-antibaryon asymmetry. Again, this stems simply from the fact that antibaryons are more matter than antimatter. Conversely, an equal number of antinucleons and nucleons does not mean there is an equal number of antistars and stars because the gravitational force between antinucleons is about one-ninth of that between nucleons and so antistar formation is suppressed relative to star formation.
\end{abstract}
When Einstein first articulated the General Theory of Relativity (GR), the "stuff" in the universe that interacted gravitationally (i.e., matter) was atoms and electromagnetic radiation\footnote{The discussion in this article will be framed in the language of particle physics so I will, for the most part, talk about photons rather than EM radiation. It is somewhat fitting, and ironic, that about the only person who believed in photons, or light-quanta, in 1919 was Albert Einstein.}  The fact that light's path was observed\cite{Edd} to be bent by the Sun as predicted by GR demonstrated that photons were just another form of mass-energy. Comparisons of 
measurements of the deflection of electromagnetic radiation by the sun to the predictions of
GR are given in \cite{will} with modern measurements using, for example, radio
waves. Any potential difference from GR (i.e., extensions
to GR or different types of metric theories) is incorporated in the parameter $\gamma$. For
GR, $\gamma$ = 1. The result as of 2004 from 87 VLBI sites using almost 2 million VLBI
observations of 541 radio sources is $\gamma -1 = (-1.7\pm 4.5)\times 10^{-4}$. 

In the 1950's Bondi\cite{bondi} introduced two types of gravitational mass -- "passive" gravitational mass (the mass that interacts with a gravitational field) and "active" gravitational mass (the mass that is a source of gravitational field). Under this distinction the Weak Equivalence Principle (WEP) then relates inertial mass to passive gravitational mass.
There is a 2023 result\cite{llr} using lunar laser ranging of the Earth-Moon distance to study the possible violation of the equivalence of active and passive gravitational mass, $m_a/m_p$. For two bodies $A$ and $B$, they define the variable $S_{A,B}$ as
$$S_{A,B}=\frac{m_{aB}}{m_{pB}}-\frac{m_{aA}}{m_{pA}}$$
which they assert plays a role analogous to the E\"otv\"os parameter in tests of WEP. For the two elements relevant for the Moon, iron and aluminum, they
find a limit of $S_{Al,Fe}<3.9\times 10^{-14}$. From hereon in I will not distinguish between passive and active gravitational mass.

The atom, of course, consists of: a nucleus and electrons that are bound electromagnetically, nucleons bound by the nuclear force, and quarks in the protons and neutrons bound by the strong force. 
If binding energy
coupled to gravity differently than particulate matter (i.e., the particles that are bound by the interaction) then one would observe a violation of WEP. This is represented in \cite{will} in the following way:
$$m_G=m_I+\sum_A\frac{\eta^AE^A}{c^2}$$
where $m_G$ is the gravitational mass, $m_I$ is the inertial mass, $E^A$ is  the  internal  energy  of  the  body  generated  by  interaction $A$, and $\eta^A$ is a dimensionless parameter that measures the strength of the violation of WEP induced by that interaction. 
The WEP violation parameter $\eta^A$ is related to the E\"otv\"os parameter, $\eta$, for objects 1 and 2 by:
$$\eta\equiv \frac{\vert a_1-a_2\vert}{(a_1+a_2)/2}=\sum_A\eta^A\left(\frac{E_1^A}{m_1c^2}-\frac{E_2^A}{m_2c^2}\right)$$
where $a$ is the measured free-fall acceleration. The E\"otv\"os parameter
has been found to be very small (less than a part in $10^{13}$) in classic torsion balance experiments\cite{WEP_tests}. The 2022 result\cite{MIC} from the space-based MICROSCOPE experiment using titanium and platinum is:
$$\eta(Ti,Pt)=[-1.5\pm 2.3(stat)\pm 1.5(syst)]\times 10^{-15}$$
The bottom line is that, to very high precision, there is no difference in the gravitational interaction between particle mass-energy and binding energy, as predicted by (actually built into) GR. That is, {\bf particle mass-energy and binding energy both act gravitationally as matter}.

\section{Antiparticles and Gravity}

In principle, the subsequent discovery of the positron, the antielectron, did little more
than add to the list of existing fundamental ”matter” particles; the proton, neutron, electron, and
photon. But the addition of antimatter raised the issue of whether it might interact differently from matter under gravitation.  
The idea of "antigravity", that matter and antimatter repel gravitationally\footnote{Reference \cite{dream} is a delightful 1898(!) letter to {\it nature} by A. Schuster. 
In this wonderfully whimsical paper, matter and antimatter 
are thought of as sources and sinks of some fluid. To quote from the paper, "These sinks would form another set of atoms, possibly equal to our own in all respects but one; they would mutually gravitate towards each other, but be repelled from the matter which we deal with on this earth." 
Further, "If there is negative electricity, why not negative gold, as yellow and valuable as our own, with the same boiling point and identical spectral lines; different only in so far
that if brought down to us it would {\it rise up into space with an acceleration of 981} [my italics]." As far as I know, this paper includes the first use of the term "antimatter" although not, obviously, related in any way to the present meaning since it predates Dirac's prediction of the existence of antimatter by some 30 years.
Many thanks to my colleague Joseph McKenna for pointing me to this paper.}, has, on the face of it, some attractive qualitative features. One of the great open questions in physics is, "Where did all the antimatter created in the Big Bang go?" But if matter and antimatter repel gravitationally then this question is moot -- there are equal amounts of matter and antimatter in the universe but matter and antimatter would separately coalesce into galaxies and antigalaxies, respectively, which would repel each other. This is completely consistent with the observed paucity of antimatter around us\footnote{Again from Schuster\cite{dream}, "The fact that we are not acquainted with such matter {\rm [i.e., antimatter]} does not prove its non-existence; for
if it ever existed on our earth, it would long have been repelled by it and expelled from it. Some day we may detect a mutual repulsion between different star groups, and obtain a sound footing for
what at present is only a random flight of the imagination."} as antimatter remnants would be repelled by our galaxy and so not reach us. Further, having a distribution of galaxies and antigalaxies which are repelling each other could naturally lead to an expanding universe.  

Antigravity was first seriously tackled theoretically in the 1950's\cite{matz}\cite{morr}\cite{schiff1}\cite{schiff2} and in those cases, and more recent efforts\cite{jent}\cite{kowitt}\cite{chard}, it was generally found to be problematic. The arguments against antigravity are scrutinized in \cite{nieto} (these counter-arguments are summarized in \cite{chardin}). Antigravity was generally enabled by assigning to antimatter a negative gravitational mass. This leads to the central question, "What mass carries the label 'antimatter' within the context of antigravity?" or, as in the title of this article, "What {\it Exactly} is Antimatter (Gravitationally Speaking)?"
That binding energy behaves as matter and not antimatter was recognized early on. As L.I. Schiff put it in 1958\cite{schiff2}, "electromagnetic and nuclear binding energy have positive gravitational mass." The Standard Model of particle physics contains three types of objects: fermions (e.g., $u$ quark, $\mu^-$ lepton, etc.), antifermions (e.g., $\bar{c}$, $\bar{\nu}_\tau$, etc.), and the intermediate vector bosons (e.g., photon, $W^-$, etc.). It seems clear that the antifermions are, by definition, "antimatter". To be as precise as possible, the mass-energy associated with antifermions is the "antimatter" in antigravity, i.e., the antimatter in the statement "antimatter-matter repel gravitationally." It follows that the mass-energy associated with fermions is "matter". We have also seen that the intermediate vector bosons (i.e., the fields which constitute "binding energy") are strongly experimentally constrained to behave gravitationally like "matter". 


\section{Villata Antigravity}

All the theoretical attempts to describe antigravity discussed so far don't explicitly account for binding energy or, equivalently, they don't have anything to say about why, for example, photons behave gravitationally as matter. A quite different approach which does include the photon was put forward by Villata\cite{vill1}(2011). Villata starts with the assumption that $CPT$ Invariance holds for GR. This is realized in GR through the transformations $dx^\mu\rightarrow -dx^\mu$ and $q\rightarrow -q$ where $q$ is the electric charge.
Applying $CPT$ to the equation of motion (the geodesic equation) for matter-matter interactions leads to the standard form:
\begin{equation}
    \frac{dx^\lambda}{d\tau^2}=-\frac{m_{(g)}}{m_{(i)}}\frac{dx^\mu}{d\tau}\Gamma^\lambda_{\mu\nu}\frac{dx^\nu}{d\tau}
\end{equation}
where he thought it "may be useful to keep the ratio
$m_{(g)}/m_{(i)}$ = 1 visible in the equation." That is, the Equivalence Principle is still enforced. For matter-antimatter interactions he finds:
\begin{equation}
    \frac{dx^\lambda}{d\tau^2}=-\frac{-m_{(g)}}{m_{(i)}}\frac{dx^\mu}{d\tau}\Gamma^\lambda_{\mu\nu}\frac{dx^\nu}{d\tau}
\end{equation}
where you can see that applying $CPT$ has introduced an extra minus sign, i.e., there is now matter-antimatter repulsion. 

Again quoting from \cite{vill1};
\begin{quote}
\begin{it}
    The minus sign assigned to the gravitational mass
in eq. (2) must not be misinterpreted. It does not mean that $m_{(g)}$ has become negative,
since, according to our assumptions, i.e. $CPT$ invariance and weak equivalence principle, all
masses are and remain positive definite. As already said, the minus sign comes from the $PT$-
oddness of either $dx^\mu$ or $\Gamma^\lambda_{\mu\nu}$. Assigning it to the mass can just be useful for not losing it when dealing with the Newtonian approximation, where four-velocities disappear, together with
their changed signs. Similarly, the Newtonian-limit field $GM/r^2$ has lost the $PT$-oddness, so
that the minus sign of an antimatter field may consequently be assigned to $M$. As a result, we
would obtain the generalized Newton law 
\begin{equation}
F(r)=-G\frac{(\pm m)(\pm M )}{r^2} = \mp G\frac{mM}{r^2}
\end{equation}
where the minus sign refers to the gravitational self-attraction of both matter and antimatter,
while the plus sign indicates the gravitational repulsion between matter and antimatter.
\end{it}    
\end{quote}
We'll come back to eq. (3) when discussing the free-fall acceleration of antihydrogen.  

The question of where photons fit into all this is addressed in Villata's response\cite{vill4} to some published critiques\cite{vill2}\cite{vill3} of his original paper. Among other things he shows via the same analysis is that the geodesic equation for photons is formally identical to that for matter particles;
$$\frac{d^2x^\lambda}{d\sigma^2}=-\Gamma^\lambda_{\mu\nu}\frac{dx^\mu}{d\sigma}\frac{dx^\nu}{d\sigma}$$
where $\sigma$ is a parameter describing the trajectory,
which for photons can't be equal to the proper
time $\tau$, since $d\tau$ = 0, and its elements have the
same ($C$)$PT$ properties as in eq. (1).

The biggest caveat, of course, is that matter-antimatter repulsion only derives from GR {\it if} $CPT$ invariance applies to the Einstein equation. The validity of this assumption was questioned in \cite{vill2} in the general sense and the applicability of $CPT$ invariance in GR is rigorously discussed in \cite{CES}. Villata justifies this assumption in the following way\cite{vill4},
\begin{quote}
    \begin{it}
        ... as it is well known, the CPT theorem is well established in flat space-time, and, even if it is not
proven in curved space-time, nobody expects that
general relativity violates it (apart from the case
of very peculiar topologies).
    \end{it}
\end{quote}

The main conclusions from Villata's analysis are; (1) it follows naturally from GR itself that matter and antimatter repel gravitationally and (2) the photon (and by extension the other intermediate vector bosons) behaves gravitationally like matter. It's nice that (2) is equivalent to saying that binding energy behaves like matter, as found experimentally through precision tests of WEP. Of course the real power of the Villata approach is that {\bf you get matter-antimatter repulsion without any changes to GR or any new interactions}. 

\section{How Much of the Antiproton is Antimatter?}


It has been established experimentally, and follows from Villata's theoretical approach, that binding energy acts gravitationally like matter so the mass of a bound system containing antifermions then has both matter and antimatter components.\footnote{To reiterate, binding energy contributes to the mass of a composite object and it has to be considered either matter or antimatter. In the case of hydrogen and antihydrogen it might be thought that the whole atom/antiatom is matter/antimatter but this leads to an inconsistency when one considers "atoms" like positronium, the bound state of a fermion (the electron) and an antifermion (the positron). The binding energy is in all cases the same stuff (electromagnetic field aka photons) and so it can't be said that it could either be matter or antimatter. Binding energy acts gravitationally as matter in this and all cases.} 
Whereas the atomic binding energy of hydrogen (13.6 eV) is many orders of magnitude smaller than the masses\footnote{I am using the particle physicist's "natural units" convention $c$ = $\hbar$ = 1 so masses are given in eV.} of the constituents -- 938.27 MeV for the proton and 511 keV for the electron -- the situation for the proton is far different.
The proton itself is a composite system composed of two up quarks and one down quark bound by the strong force. The quark masses\cite{pdg} are $m_u$ = $2.16^{+0.49}_{-0.28}$ MeV and $m_d$ = $4.67^{+0.48}_{-0.17}$ MeV and so contribute only $\sim 1$\% to the proton's mass. Or, as put by Frank Wilczek\cite{wilczek},
\begin{quote}
\begin{it}
Most of the mass of standard
matter, by far, arises dynamically, from back-reaction of the color gluon fields of quantum chromodynamics (QCD). Additional quantitatively small, though physically crucial, contributions come
from the intrinsic masses of elementary quanta (electrons and quarks).

\end{it}
\end{quote}
That is, gluonic binding energy is a major component of the mass of the proton. 
There is also the contribution to the proton mass due to the kinetic energy of the quarks. Confining quarks to a region the size of the proton gives, from Uncertainty Principle arguments, an expected quark kinetic energy contribution of around 300 MeV. These are relevant considerations for antigravity because, due to the $CPT$ Invariance of QCD, any conclusions about the contributions of quarks to the mass of the proton apply equally to antiquarks and the antiproton. 

\subsection{Theoretical Results on the Mass Structure of the Proton}

The key quantity in GR is not mass per se (since "massless" objects like the photon are affected by gravity, i.e., follow geodesics) but the Energy-Momentum Tensor (EMT). Ji found in 1995\cite{Ji1} that the total QCD EMT for the proton can be decomposed as; 
\begin{equation}
T^{\mu\nu}=T_q^{\mu\nu}+T_g^{\mu\nu}+\hat{T}^{\mu\nu}
\end{equation}
where the last term is the "trace term" (a completely QFT effect). $T^{\mu\nu}$ and $\hat{T}^{\mu\nu}$ are scheme and scale ($\mu$) independent as is the sum $T_q^{\mu\nu}(\mu)+T_g^{\mu\nu}(\mu)$. The mass $M$ is then
$$M=\int d^3\vec{x}\langle T_q^{00}(\mu)+T_g^{00}(\mu)+\hat{T}^{00}\rangle$$
The total Hamiltonian is $$H_{\rm QCD}=H_q(\mu)+H_g(\mu)+H_a$$ where, for example 
$$H_g(\mu)\equiv\int d^3\vec{x}T_g^{00}(\vec{x},\mu)
=\frac{1}{2}\int d^3\vec{x}\left(\vec{E}^2+\vec{B}^2\right)_{\rm R}(\vec{x},\mu)$$
$H_{\rm QCD}$ and $H_a$ are scheme and scale independent as is the sum $H_q(\mu)+H_g(\mu)$. Finally
\begin{eqnarray*}
    M&=&M_q+M_g+M_a\\
    &=&\frac{3}{4}M(\mu)+\frac{3}{4}M(\mu)+\frac{1}{4}M
\end{eqnarray*}
As stated in \cite{Ji2};
\begin{quote}
    \begin{it}
  If the quark masses are non-zero, one ends up with a decomposition with four terms, each of which are related to experimental observables and calculable in lattice QCD. The largest error comes from the
anomalous energy calculation, and the state-of-art
verification\cite{He} of the above sum rule is at 10\% level.      
    \end{it}
\end{quote}
There is still some disagreement in the theoretical community as to the robustness of this decomposition (see, for example, \cite{metz}\cite{lorce}(2021)). These criticisms are confronted in \cite{Ji2}(2021). 
However, the exact details aren't so important here since there is no disagreement that there is a large contribution to the proton mass from gluons (i.e., binding energy) although a hefty uncertainty must be placed on the numerical value of this contribution.

A full lattice QCD calculation\cite{yang}(2018) gave the following 4-term (\'a la Ji) proton mass decomposition: quark condensate ($\sim$9\%), quark kinetic energy ($\sim$32\%), gluonic field strength ($\sim$37\%), and anomalous gluonic contribution ($\sim$23\%). The result was given for a specific choice of scale, $\mu$ = 2 GeV. The sum of quark kinetic energy and gluon field strength contributions is unaffected by the choice of scale but the relative sizes of them is scale dependent. From this we see that the bulk of the proton mass ($\sim$68\%) is due to the gluon (colour) field. 

Lattice QCD calculations\cite{bhatt}(2023) are further pinpointing the roles played by the up and down quarks within the proton.
The authors claim that their "calculations show that the up quark is more symmetrically distributed and spread over a smaller distance than the down quark. These differences imply that up and down quarks may make different contributions to the fundamental properties and structure of the proton, including its internal energy and spin." 

\subsection{Experimental Results on the Mass Structure of the Proton}
Recent measurements are not only consistent with the theoretical analysis but show that the proton structure is even richer than imagined. 
The relevant measurable quantities are the gravitational form factors (GFF) which are related to the QCD EMT\cite{Ji2}\cite{shan}. These can be accessed through measurements of near-threshold $J/\psi$ photoproduction. The end-product is the mass radius of the proton, which is derived in a model-dependent way from the differential cross-section. Again there is some contention as to how precise the results are since things like open charm production, for example, can muddy the waters\cite{winney}. 

Even given these uncertainties a pretty clear picture emerges from the experimental data of how mass is distributed {\it within} the proton. Two groups published results on $J/\psi$ photoproduction in 2023 -- GlueX\cite{gluex} and $J/\psi$-007\cite{core}. Gluex reports mass radius values for three photon energies, specifically $\sqrt{\langle r^2_m\rangle}$ = $0.619\pm 0.094$ fm, $0.464\pm 0.024$ fm, and $0.521\pm 0.020$ fm for $E_\gamma$ = 8.93, 9.86, and 10.82 GeV, respectively. $J/\psi$-007 find using so-called holographic QCD a mass radius of $0.755\pm 0.035$ fm and a scalar radius of $1.069\pm 0.056$ fm. These values are in good agreement with results extracted from lattice QCD\cite{pefkou}\cite{mamo}. This is to be compared to the charge radius value of the proton of around 0.84 fm. The $J/\psi$-007 group interpret their result in the following way:
\begin{quote}
    \begin{it}
        We therefore propose that the structure consists of three distinct regions. An inner core, dominated by the tensor gluonic field structure, provides most of the mass of the proton. The charge radius, determined by the relativistic motion of the quarks, extends beyond this inner core. The entirety of the proton is enveloped in a confining scalar gluon density, extending well beyond the charge radius. 
    \end{it}
\end{quote}

\subsection{The Mass Structure of the Antiproton}
The theoretical and experimental results give a clear and consistent picture as to how mass is distributed within the proton. 
The WEP tests described earlier require that all of the terms in the proton EMT, eq. (4), must be considered "matter" since all of the proton mass is matter. Due to the $CPT$ invariance of QCD, all of the EMT terms for the antiproton have the same numerical value with the only difference being that 
the quark kinetic energy term becomes the antiquark kinetic energy term and should carry the label antimatter. The quark condensate EMT term, essentially the "sea" quarks or virtual $q\bar{q}$ pairs generated by the colour field, is the same for the antiproton and proton and thus operates under gravity like matter. 

To compare antigravity predictions to the experimental findings we need some estimate of the uncertainty in the antiquark contributions to the antiproton mass. The Lattice QCD result quoted above for the quark kinetic energy term was $(32\pm 4~({\rm stat})\pm 4~({\rm syst}))$\%. Analytical results for this term using different schemes\cite{metz} give a consistently lower value. 
Hence, to be conservative, I'll assign a 30\% uncertainty to the lower bound so that the antimatter fraction of the antiproton mass is then $(33^{+6}_{-10})\%$ -- 32\% from antiquark kinetic energy and 1\% from antiquark masses. 
The uncertainty here may actually be underestimated but debating the exact value should not distract from the central message of this paper -- that the antimatter fraction of the antiproton (and, hence, also the antineutron) is certainly less than 50\% and so, with regard to gravitational interactions, an antiatom is actually more matter than antimatter.  

\section{Implications for the Free-fall Acceleration of Antihydrogen}

The fact that an antinucleon would not simply be repelled by the Earth even in the antigravity scenario was understood as far back as 1958\cite{morr} when P. Morrison wrote:
\begin{quote}
    \begin{it}
        Since there is no reason to expect that electromagnetic or pi-mesic energy is absent from the total rest energy of a nucleon, it follows that some unknown but probably not major part of the gravitational mass of the antinucleon arises from these sources, and would not be repelled by the earth's field. Therefore, an antinucleon need not show the acceleration $\vert g\vert$ in its free rise. It is clear that there is no Lorentz covariant way to describe the separation of the mass of a particle into these two kinds. 
    \end{it}
\end{quote}
It was true then and is true now that electromagnetic and "pi-mesic" (i.e., nuclear) binding energy are not a major part of the mass of an antinucleon (they're actually completely insignificant). 
Morrison was writing before the discovery of quarks and QCD and so had no way of knowing that, in fact, the majority of the mass of an antinucleon comes from gluonic binding energy, making his observation even more prescient now than it was then.
The Lorentz covariant way to describe the separation of mass into these "two kinds" is quite natural within the framework of the QCD EMT partitioning given by eq. (4). 

There were two papers\cite{lykken}(2008)\cite{alves}(2009) 
that discussed various measurements which indirectly set limits on $\Delta\equiv \vert g_H-g_{\bar{H}}\vert/g$. In \cite{lykken} they explicitly comment on the fact that the rest masses of the antiquarks constitute only around $\sim 1$\% of the antiproton mass and so consider this a kind of hard upper limit on $\Delta$. This paper came out before the Lattice QCD calculation\cite{yang} 
showing that $\sim 32$\% of the proton mass is due to the quark kinetic energy.
In \cite{alves} they state, 
\begin{quote}
    \begin{it}
        The essential point is that nuclei and atoms are
composite states. Although one can make a distinction
between matter and antimatter at the level of quarks
and electrons, that distinction is blurred when one
considers bound states like nuclei and atoms. And
because antimatter plays a quantifiable role in the
physics of nuclei and atoms by contributing to their
inertial masses, precision E\"otv\"os experiments utilizing
matter continue to be relevant when considering the
possibility of gravitational asymmetry between matter
and antimatter.
    \end{it}
\end{quote}
The claim here is that the antimatter in virtual fermion-antifermion loops contribute to the inertial mass of nuclei and atoms.
However, virtual antifermions do not have a "quantifiable role" in the antimatter content of nucleons. The "sea" quarks and antiquarks are in the quark condensate EMT term, a term that actually behaves as matter under gravitation. It seems plausible that it is also true for the EMT term for the $e^+e^-$ pairs in, for example, the Lamb Shift calculation. That is, again, the WEP tests constrain the atomic and nuclear virtual loops to act gravitationally like matter. So they wouldn't have any affect on QED calculations given in \cite{alves} and wouldn't therefore contribute to any bound on $\Delta$.

The bounds on $\Delta$ are also considered in \cite{blas}(2017) where various arguments are given for how they could be circumvented to some degree. 
This is another paper written before the Lattice QCD calculation was published so when confronting the "1\%" argument they write,
\begin{quote}
    \begin{it}
   This line of reasoning is correct, but the argument based on the energy in the nucleon
in the form of gluons (or partons) is not conclusive. The reason is that it is not very
well formulated since we do not know of any complete theory that simply differentiates
between the different types of energy present in a nucleon.     
    \end{it}
\end{quote}
The Lattice QCD calculation does, in fact, differentiate between the different types of energy present in a nucleon, as described in Section 3.1. 

\section{Estimation of the Free-fall Acceleration of Antihydrogen}

To predict the free-fall acceleration for antihydrogen I will use eq (3).
 To implement this equation for a composite system requires disentangling which components of the mass-energy of an object are antimatter and which are matter.
It is stated in \cite{khar} (after a rigorous derivation of Newton's Law of Gravitation from GR):
\begin{quote}
    \begin{it}
        The purpose of reviewing this textbook derivation was to show that in the weak gravitational field, non-relativistic, limit of gravity the distribution of mass and the distribution of the trace of the EMT are identical.
    \end{it}
\end{quote}
Recall that the matter and antimatter fractions of the antiproton mass are derived from the energy-momentum tensor.
Taking $m$ and $\bar{m}$ as the matter and antimatter components of an object having total mass $M$ (recall that binding energy is part of $m$) we have
$$M=m+\bar{m}=M(f+\bar{f})$$
where $f$ and $\bar{f}$ are the fractions of the object's mass which are matter and antimatter, respectively.

I will rewrite eq (3) in the form:
\begin{equation}
F=\frac{-M_1G_{12}M_2}{r_{12}^2}
\end{equation}
where
$f_1G_{12}f_2=+Gf_1f_2$ and $\bar{f}_1G_{12}\bar{f}_2=+G\bar{f}_1\bar{f}_2$ while $\bar{f}_1G_{12}f_2=-G\bar{f}_1f_2$ and $f_1G_{12}\bar{f}_2=-Gf_1\bar{f}_2$ (i.e., matter attracts matter and antimatter attracts antimatter while matter and antimatter repel).

Now, with $M_1=M_1(f_1+\bar{f}_1)$ and $M_2=M_2(f_2+\bar{f}_2)$ we then have:
\begin{eqnarray*}
    F&=&\frac{-M_1G_{12}M_2}{r_{12}^2}\\
    &=&\frac{-M_1(f_1+\bar{f}_1)G_{12}M_2(f_2+\bar{f}_2)}{r_{12}^2}\\
    &=&\frac{-M_1M_2(f_1G_{12}f_2+\bar{f}_1G_{12}\bar{f}_2+f_1G_{12}\bar{f}_2+\bar{f}_1G_{12}f_2)}{r_{12}^2}\\
    &=&\frac{-GM_1M_2(f_1f_2+\bar{f}_1\bar{f}_2-f_1\bar{f}_2-\bar{f}_1f_2)}{r_{12}^2}\\
    &=&\frac{-GM_1M_2\left((1-\bar{f}_1)(1-\bar{f}_2)+\bar{f}_1\bar{f}_2-(1-\bar{f}_1)\bar{f}_2
    -\bar{f}_1(1-\bar{f}_2)\right)}{r_{12}^2}
\end{eqnarray*}
I replaced $f_{1,2}$ with $(1-\bar{f}_{1,2})$ because the matter and antimatter fractions are anticorrelated and so, in order to incorporate the uncertainties from the Lattice QCD calculations, it only makes sense to include in the final expression either the matter fraction or the antimatter fraction. Finally we arrive at:
\begin{equation}
    F=\frac{-GM_1M_2\left(1-2(\bar{f}_1+\bar{f}_2)+4\bar{f}_1\bar{f}_2\right)}{r_{12}^2}
\end{equation}
For the case of antihydrogen falling in the gravitational field of the Earth, eq. (6) becomes particularly simple since then $\bar{f}_1=0$. Equating $M_1=M_E$ and $r_{12}=R_E$ we then have:
$$  F=\frac{-GM_EM_{\bar{H}}(1-2\bar{f}_{\bar{H}})}{R_E^2} $$ 
and it follows that the free-fall acceleration of antihydrogen is:
$$a_{\bar{H}}=(1-2\bar{f}_{\bar{H}})g$$
Note that if $\bar{f}_{\bar{H}}$ were equal to 1 (i.e., antihydrogen was completely composed of antimatter) then $a_{\bar{H}}=-g$ and it would be accurate to say that the ALPHA-g result precluded antimatter falling up. The contribution to the antimatter fraction of antihydrogen due to the positron (and its kinetic energy) is negligible so effectively $\bar{f}_{\bar{H}}=\bar{f}_{\bar{p}}$. As we have seen, $\bar{f}_{\bar{p}}\ne 1$ but is closer to one-third and so it is more accurate to say that ALPHA-g didn't show that antimatter doesn't fall up but, in fact, that an antiatom (and, so, an antiapple!) does not fall up.

What, then, would be the free-fall acceleration of antihydrogen given our present knowledge of the mass structure of the antiproton?  We have from Section 3.1 that $\bar{f}_{\bar{p}}\approx 0.33^{+0.06}_{-0.10}$ so,
in the antigravity scenario (matter and antimatter repel), the free-fall acceleration of antihydrogen would be $a_{\bar{H}}=(0.33^{+0.23}_{-0.11})g$. Just simply adding the uncertainties in quadrature (a fuller analysis is required to get the error on the difference right) gives the  difference between the ALPHA-g measurement and the antigravity prediction as $(0.42\pm 0.23)g$. The antigravity prediction is disfavoured but not ruled out by the ALPHA-g measurement. If ALPHA-g can reduce the experimental uncertainty to less than $\sim 5\%$ or so then antigravity could be definitively ruled out (or verified!).  

\subsection{Free-fall Acceleration of Muonium and of Positronium}
There is a proposal\cite{mage} to measure the free-fall acceleration of muonium, the bound state of an electron and an antimuon. The binding energy of muonium (essentially that of hydrogen, i.e., 13.6 eV) and the kinetic energy of the electron are negligible compared to the $\mu^+$ mass of 105.66 MeV = 206.77$m_e$. So for muonium:
$$\bar{f}_{\rm Mu}= \frac{m_\mu +K_\mu}{m_\mu+K_\mu+m_e+K_e+E_B}\approx \frac{206.77m_e}{207.66m_e+me}=0.995$$
and we would therefore expect:
$$a_{\rm Mu}=(1-2\bar{f}_{\rm Mu})g=(1-2(0.995))g=-0.99g$$
That is, muonium would indeed "fall up" with an acceleration very close to $g$. 

The binding energy for positronium is, at half that of hydrogen (i.e., $\sim$ 7 eV), negligible compared to the mass of the electron/positron. Therefore, in principle, positronium is an almost completely equal mix of matter and antimatter and should basically hover in the gravitational field of the Earth. 
Unfortunately it has a lifetime of only 142 ns so it is presumably not feasible to make a measurement of the free-fall acceleration of positronium.


\section{Cosmological Implications of Binding Energy as Matter}
 
The precise definition given here of antimatter as antifermion mass-energy and matter as fermion and intermediate vector boson mass-energy results in a much different and richer universe in the antigravity scenario than for the naive version containing "equal amounts of matter and antimatter"\footnote{I can't help but quote from Schuster\cite{dream} once again, this time on his antimatter fluid universe. "Whether
such thoughts are ridiculed as the inspirations of madness, or allowed to be the serious possibilities of a future science, they add renewed interest to the careful examination of the incipient worlds which our telescopes have revealed to us. Astronomy, the oldest and yet most juvenile of sciences, may still have some surprises in store. May antimatter be commended to its care! But I must stop -- the holidays are nearing their end -- the
British Association is looming in the distance; we must return to sober science, and dreams must go to sleep till
next year."}. It is not necessarily {\it observably} different from the one we inhabit but its evolution and structure will almost certainly be. For starters, the matter-antimatter asymmetry is there right from the instant of the Big Bang. That is, 
there are indeed equal amounts of fermions (matter) and antifermions (antimatter) created at the Big Bang but also some number of intermediate vector bosons (or the $X$ and $Y$ bosons from GUT theories) which act gravitationally as matter. So right from the start there is more matter than antimatter in the universe. Presumably matter and antimatter would congregate into separate "islands" which would repel each other 
and this separation of antifermion islands from the rest would exist through inflation ($10^{-33}$ to $10^{-32}$ seconds after the Big Bang). 

A really interesting thing happens in the hadronization phase (approximately a microsecond after the Big Bang). The nucleons form as usual but the antimatter dominated regions go through a kind of phase change. That is, the antifermions are indeed pure antimatter but antinucleons are roughly two-thirds matter so now even the previous pockets of antimatter become dominantly matter. The antinucleons would combine with positrons to make antihydrogen and antihelium\footnote{The STAR Collaboration at RHIC measured the strength of the interaction between antiprotons\cite{STAR}(2015) and found it to be consistent with that between protons. And so it seems reasonablel to assume, just as $CPT$ Invariance demands, that antihelium has the exact same properties as helium.}, just as the nucleons would atomize. But the antihydrogen and antihelium wouldn't form antistars and antigalaxies at anywhere near the same rate as for the matter sector because 
the gravitational attraction between antiatoms is roughly one-ninth of that between atoms. This follows from eq. (6) where we get the force between two antinucleons as:
\begin{eqnarray*}
        F&=&\frac{-GM_1M_2\left(1-2(\bar{f}_1+\bar{f}_2)+4\bar{f}_1\bar{f}_2\right)}{r_{12}^2}\\
        &\approx&\frac{-GM_{\bar{N}}^2}{r^2}\left[1-2\left(\frac{1}{3}+\frac{1}{3}\right)+4\left(\frac{1}{3}\right)\left(\frac{1}{3}\right) \right]\\
        &=&\frac{-GM_N^2}{r^2}\left[\frac{1}{9}\right]
\end{eqnarray*}
with $M_1=M_2=M_{\bar{N}}=M_N=M_p$ (or $M_n$) and I have assumed the antimatter fraction is the same for antiprotons and antineutrons (it should be very close to the same). An analogous calculation (the one done in Section 5 for antihydrogen free-fall) gives that the nulceon-antinucleon gravitational force is about one-third that of the gravitational force between nucleons. 

To fully understand the implications of antigravity requires simulations of cosmological evolution that incorporate what constitutes antimatter and what constitutes matter and how nucleons and antinucleons interact. However, even without complete simulations, various general features of the universe in this picture emerge: 
\begin{enumerate}
\item Though there are as many antinucleons\footnote{I will continue to say antinucleon and nucleon although by this point in the evolution of the universe they are dominantly bound states of antihydrogen and hydrogen with some antihelium and helium thrown in.} as nucleons in the antigravity universe, the fact that $F_G(\bar{N}\bar{N})\approx F_G(NN)/9$ means that antistar formation will be suppressed relative to star formation. It might be even further suppressed because, though the antinucleons would be initially spatially isolated from the nucleons, $F_G(N\bar{N})\approx F_G(NN)/3$ and so the antinucleonic cloud is also attracted to the surrounding matter. This would seemingly lead to regions containing diffuse clouds of antinucleons with some antistars thrown in. It's also possible that galaxies could be enmeshed in a much larger, diffuse cloud of antinucleons or that galaxies would contain some small numbers of either antistars or clouds of antinucleons. Again, without a proper calculation, it's hard to know if the 2021 observation\cite{antistar} of some fourteen antistars in our galaxy or the purported observation of eight antihelium events by AMS-02\cite{AMS}(2018) is consistent with this. 
\item Since antiquarks are bound into matter-dominated hadrons, the only purely antimatter objects in the universe are antineutrinos and positrons. These would be repelled by everything else in the universe (since even the antinucleons are dominantly matter) and so presumably coalesce into streams or islands. This could be tested if one could measure the relative number of neutrinos to antineutrinos in the Cosmic Neutrino Background (the C$\nu$B). In the antigravity scenario we should observe very few if any C$\nu$B antineutrinos locally. It's not clear if these regions of pure antimatter are enough to account for the accelerated expansion of the universe as described in, for example, \cite{vdark}.
\item The regions with a high density of antinucleons will act as gravitational lenses but with a longer focal length than for regular matter since $F_G(\gamma\bar{N})\approx F_G(\gamma N)/3$. Further, regions with high densities of antineutrinos and/or positrons would gravitationally defocus light. This considerably complicates the interpretation of lensing data in terms of the amount of mass there is in the universe.
\end{enumerate}
I will elaborate on these issues and others (e.g., antigravity and black holes) in more detail in a subsequent article.
\section{Conclusions}

The main goal of this paper was to clarify what is meant by the term "antimatter", to make it clear that there is a distinction that needs to be made when talking about antigravity, namely, the term "antimatter" is not synonymous with "antiparticle".
Tests of the Weak Equivalence Principle show that binding energy acts gravitationally like matter while Lattice QCD calculations find that only about one-third of the proton's mass is quark mass-energy. QCD is $CPT$ invariant and so it follows that the same fraction of the antiproton mass is due to the mass-energy of the antiquarks. That is, two-thirds of the mass of the antiproton (gluonic binding energy) carries the label "matter" and one-third (antiquark mass-energy) is "antimatter".  This has a number of implications for various observables in the antigravity scenario, i.e., a universe where matter and antimatter repel gravitationally. In this scenario, one would expect the free-fall acceleration of antihydrogen to be $a_{\bar{H}}=(0.33^{+0.23}_{-0.22})g$. This is to be compared to the recent ALPHA-g result of $a_{\bar{H}}=(0.75\pm 0.13~({\rm stat.+syst.})\pm 0.16~({\rm simulation}))g$. Further, the universe would evolve much differently than in the current paradigm. For example, it follows naturally that there is a matter-antimatter asymmetry in the universe (right from the instant of the Big Bang, in fact) but not a baryon-antibaryon asymmetry (or, more directly, not a quark-antiquark asymmetry). Again, this stems simply from the fact that antibaryons are more matter than antimatter. However, an equal number of antinucleons and nucleons does not mean there is an equal number of antistars and stars because the gravitational force between antinucleons is about one-ninth of that between nucleons. Hence, a much smaller but non-zero number of antistars compared to stars is not an argument against antigravity but is possible evidence for a it.

\section{Acknowledgements}
I gratefully acknowledge the countless conversations about antimatter and gravity with many of my ALPHA-g colleagues, in particular Makoto Fujiwara and Andrea Capra. I also appreciate the efforts of Randy Lewis to educate me about the subtleties of the Lattice QCD calculation of the proton mass composition. None of the above are to be blamed for anything I have written here.

\end{document}